# Continuous Magnetophoretic Separation of Blood Cells from Plasma at the Microscale

E. P. Furlani

*Institute for Lasers, Photonics and Biophotonics,
University at Buffalo (SUNY), Buffalo, NY, 14260*

## Abstract

We present a method for the direct and continuous separation of red and white blood cells from plasma at the microscale. The method is implemented in a microfluidic system with magnetic functionality. The fluidic structure within the microsystem consists of an inlet and a single microfluidic channel with multiple outlets. The magnetic functionality is provided by an array of integrated soft-magnetic elements that are embedded transverse and adjacent to the microchannel. The elements are magnetized using an external field, and once magnetized they produce a magnetic force on blood cells as they flow through the microchannel. In whole blood, white blood cells (WBCs) behave as diamagnetic microparticles, while red blood cells (RBCs) exhibit diamagnetic or paramagnetic behavior depending on the oxygenation of their hemoglobin. We study the motion of blood cells through the microchannel using a mathematical model that takes into account the magnetic, fluidic and gravitational forces on the cells. We use the model to study blood cell separation, and our analysis indicates that the microsystem is capable of separating WBC-rich plasma, deoxygenated RBC-rich plasma and cell-depleted plasma into respective outlets.

Corresponding Author E. P. Furlani  Email: efurlani@buffalo.edu

# I. Introduction

Microfluidic systems with magnetic functionality are finding increasing use in bioapplications that involve the immobilization and separation of biomaterials such as cells, proteins, antigens and DNA.[1] In most such applications, the biomaterial needs to be magnetically "labeled" with magnetic micro/nano-particles in order to provide sufficient coupling to an applied field to enable manipulation. Direct manipulation (without labeling) is not usually practical as the intrinsic magnetic susceptibility of most biomaterials is so small that they do not efficiently couple to a field. However, there are exceptions such as blood cells, which can be manipulated with practical magnetic fields.[2-8] Furthermore, in venous (deoxygenated) blood, white blood cells (WBCs) behave as parametric microparticles, while deoxygenated red blood cells (RBCs) exhibit diamagnetic behavior.[9-10] Thus, these two types of cells move in opposite directions in an applied field, and therefore in principle they can be magnetically separated from one another.

To date, relatively few systems have been developed that magnetically separate unlabeled blood cells, despite a substantial need for fast, accurate, and inexpensive blood cell analysis.[11] One such system has been developed by Takayasu et al.[10] In this system, the blood flows continuously through a flexible rectangular conduit (3.5 mm x 1 mm) that is wrapped in the form of a helix with an overall diameter and length of 80 mm and 100 mm, respectively. The magnetic force is provided by a magnetic wire (1mm diameter) that is wrapped along the length of the conduit in close proximity to it. The wire is magnetized by placing the conduit/wire assembly along the axis of a superconducting solenoidal electromagnet. This system separates deoxygenated blood into three outlets that contain WBC-rich plasma, deoxygenated RBC-rich plasma and cell-depleted plasma, respectively.

A much smaller microscale system has been developed by Han and Frazier.[12] In this system there are two identical and parallel flow channels, which are 30 mm long and have a height and width of 50 μm and 150 μm, respectively. The magnetic force is provided by an integrated rectangular (50 μm x 120 μm) soft-magnetic bar that runs the length of the channels (between them), and is magnetized by a permanent magnet. This system separates deoxygenated blood into three outlets; two that contain deoxygenated RBC-rich plasma, and a third that contains WBC-rich plasma.

In this paper we introduce a magnetically functional microfluidic system for the continuous and direct separation of blood cells from plasma. The fluidic structure in the microsystem consists of an inlet that feeds a single microfluidic channel with three outlets (Fig. 1). The magnetic functionality is provided by an array of integrated soft-magnetic elements, which are embedded transverse and adjacent to the microchannel. A bias field is used to magnetize the elements, and once magnetized these elements produce a magnetic force on blood cells as they flow through the microchannel. Specifically, the magnetized elements repel WBCs and attract deoxygenated RBCs (Fig. 1c). The microsystem is oriented with the microchannel (blood flow) aligned with the gravitational force, and the cells are separated perpendicular to the flow. The reason for this is because the fluidic and gravitational forces are greater than the magnetic force over extended regions of the flow channel, and therefore they need to be orthogonal to the direction of cell separation so as not to resist it. In this paper we study the transport of blood cells through the microsystem. We model blood cell separation taking into account magnetic, fluidic and gravitational forces. Our analysis indicates that the microsystem is capable of separating WBC-rich plasma, deoxygenated RBC-rich plasma and cell-depleted plasma into respective outlets.

The cell/plasma separation method presented here has advantages over competing techniques such as centrifuging, or magnetic activated cell sorting (MACS), which involves magnetic labeling. Specifically, small samples of blood can be processed within minutes with the cells in their native state. Furthermore, the microsystem is compact and efficient; i.e., energy is needed to sustain blood flow through the microchannel, but the cell/plasma separation itself consumes no energy. Thus, the microsystem is highly portable and holds potential for low cost point-of-service medical diagnostic applications. It is also ideal for filtering weakly magnetic micro/nano-particles from a transport fluid.

Lastly, we note that this paper represents an extension of earlier work.[13] Specifically, the microsystem described here is an enhancement of one introduced in a previous paper, which was limited to the separation of WBCs from deoxygenated RBCs, with no ability to extract cell-depleted plasma. Similarly, a model for predicting cell transport through the microsystem was developed earlier, but its application was limited to a proof-of-concept study. Here, we use the model for parametric analysis to study the cell separation performance of the microsystem as a function of its magnetic structure.

## II. Theory

In this section we briefly review a model for predicting blood cell transport through the microsystem; a detailed description of the model can be found in a previous paper.[13] The model takes into account the dominant forces on the cells and is based on the following simplifying assumptions: the blood cells are treated as noninteracting rigid microspheres; the blood flow in the microchannel is laminar; the impact of particle motion on the flow field is ignored; a fixed value is used for the blood viscosity (complex blood rheology is ignored); and wall effects are ignored when determining the fluidic force. While this model is not suitable for a rigorous analysis of cell transport, we use it to assess the feasibility of cell separation based on the interplay of the dominant forces on the cells. A rigorous analysis of cell transport would require a complicated, fully-coupled structural/fluidic numerical analysis that would account for the interactions of numerous structurally deformable cells, and this is beyond the scope of the present work.

We predict the motion of blood cells using Newton's law,

$$m_c \frac{d\mathbf{v}_c}{dt} = \mathbf{F}_m + \mathbf{F}_f + \mathbf{F}_g, \qquad (1)$$

where $m_c$ and $\mathbf{v}_c$ are the mass and velocity of the cell, and $\mathbf{F}_m$, $\mathbf{F}_f$, and $\mathbf{F}_g$ are the magnetic, fluidic and gravitational force (including buoyancy), respectively. The magnetic force is obtained using an "effective" dipole moment approach and is given by[14-16]

$$\mathbf{F}_m = \mu_0 V_c (\chi_c - \chi_f)(\mathbf{H}_a \bullet \nabla)\mathbf{H}_a, \qquad (2)$$

where $\chi_c$ and $V_c$ are the susceptibility and volume of the cell, and $\chi_f$ is the susceptibility of the transport fluid (in this case plasma). $\mathbf{H}_a$ is the applied magnetic field at the center of the cell, and $\mu_0 = 4\pi \times 10^{-7}$ H/m is the permeability of free space. The fluidic force is based on Stokes' law for the drag on a sphere in a viscous fluid,

$$\mathbf{F}_f = -6\pi\eta R_{c,hyd}(\mathbf{v}_c - \mathbf{v}_f), \qquad (3)$$

where $R_{c,hyd}$ is the effective hydrodynamic radius of the cell, and $\eta$ and $\mathbf{v}_f$ are the viscosity and the velocity of the fluid, respectively. The gravitational force is given by

$$\mathbf{F}_g = V_c(\rho_c - \rho_f)g\,\hat{\mathbf{x}}, \qquad (4)$$

where $\rho_c$ and $\rho_f$ are the densities of the cell and fluid, respectively ($g = 9.8$ m/s$^2$). Note that gravity acts in the + x direction, parallel to the flow (see Figs. 1c and 2b).

**A The Magnetic force**

The magnetic force depends on the field in the microchannel and the coupling of the cells to this field. The magnetic field is a superposition of the bias field $\mathbf{H}_{bias}$ and the field $\mathbf{H}_e$ due to the array of magnetized elements,

$$\begin{aligned}\mathbf{H}_a &= \mathbf{H}_{bias} + \mathbf{H}_e \\ &= H_{e,x}\hat{\mathbf{x}} + \left(H_{bias,y} + H_{e,y}\right)\hat{\mathbf{y}}.\end{aligned} \qquad (5)$$

These fields are not independent as $\mathbf{H}_{bias}$ induces $\mathbf{H}_e$. Thus, to predict the total field in the microchannel we need models for the bias field, the magnetization of the elements, and the field of the magnetized elements. Our model for the bias field is based on an analytical analysis of the field due to a rectangular block magnet[17] (pp 211-216 of reference 18). We use this model to optimize the magnitude and uniformity of the bias field across the microsystem. Once $\mathbf{H}_{bias}$ is known we use a linear magnetization model with saturation to predict the magnetization $\mathbf{M}_e$ of the soft-magnetic elements.[13] Specifically, we assume that the elements are identical and noninteracting (i.e., the field of one does not affect the magnetization of another), and obtain

$$\mathbf{M}_e = \begin{cases} \dfrac{\mathrm{H}_{bias}}{\mathrm{N}_d} & \mathrm{H}_{bias} < \mathrm{N}_d \mathrm{M}_{es} \\ \mathrm{M}_{es} & \mathrm{H}_{bias} \geq \mathrm{N}_d \mathrm{M}_{es} \end{cases}, \qquad (6)$$

where $\mathrm{N}_d$ is the demagnetization factor of the element, which is geometry dependent, and $\mathrm{M}_{es}$ is its saturation magnetization. The demagnetization factor for a highly permeable ($\chi_e \approx \infty$) long rectangular element of width 2w and height 2h that is magnetized parallel to its height can be obtained using analytical formulas (see Fig. 2b). Specifically, both the demagnetization factor $\mathrm{N}_d$ and the aspect ratio of the element $p = \dfrac{h}{w}$ can be defined parametrically as a function of a variable $k$ over the domain $0 < k < 1$ as follows:[19]

$$\mathrm{N}_d = \frac{4}{\pi} \frac{\left[\mathrm{E}(k) - k'^2 K(k)\right]\left[\mathrm{E}(k') - k^2 K(k')\right]}{k'^2}, \qquad (7)$$

$$\frac{h}{w} = \frac{\mathrm{E}(k') - k^2 K(k')}{\mathrm{E}(k) - k'^2 K(k)}, \qquad (8)$$

where $k' = \sqrt{1-k^2}$, and $K(k)$ and $E(k)$ are the complete elliptic integrals of the first and second kind, respectively,

$$K(k) = \int_0^{\frac{\pi}{2}} \frac{1}{\sqrt{1-k^2 \sin^2(\phi)}} d\phi, \qquad E(k) = \int_0^{\frac{\pi}{2}} \sqrt{1-k^2 \sin^2(\phi)} \, d\phi. \qquad (9)$$

To determine the magnetization of the elements, we first use Eqs. (7) and (8) to obtain $N_d$ for a give aspect ratio $p$ (see p 191, Table A.2 in reference 19), and then we substitute this value into Eq. (6) to get $M_e$.

Once $\mathbf{M}_e$ is known, $\mathbf{H}_e$ is easily determined. Specifically, the field solution for a long rectangular element of width 2w and height 2h that is centered with respect to the origin in the x-y plane, and magnetized parallel to its height (along the y-axis as shown in Fig. 2b) is well known (pp 210-211 in reference 18). The field components are

$$H_{ex}^{(0)}(x,y) = \frac{M_e}{4\pi} \left\{ \ln\left[\frac{(x+w)^2 + (y-h)^2}{(x+w)^2 + (y+h)^2}\right] - \ln\left[\frac{(x-w)^2 + (y-h)^2}{(x-w)^2 + (y+h)^2}\right] \right\}, \qquad (10)$$

and

$$H_{ey}^{(0)}(x,y) = \frac{M_e}{2\pi} \left\{ \tan^{-1}\left[\frac{2h(x+w)}{(x+w)^2 + y^2 - h^2}\right] - \tan^{-1}\left[\frac{2h(x-w)}{(x-w)^2 + y^2 - h^2}\right] \right\}. \qquad (11)$$

In these equations, $M_e$ is determined using Eq. (6).

The field due to the entire array of elements is obtained via superposition. Let $N_e$ denote the number of elements, and let the first element be centered with respect to the origin in the x-y plane. All other elements are positioned along the x-axis as shown in Fig. 2b. We identify the elements using the index n = (0,1,2,3,4, …, $N_e$-1). The field components due to the first element (n = 0) are given by Eqs. (10), (11). The n'th element is centered at $x = s_n$, and its field and force components can be written in terms of the 0'th components as follows:

$$H_{ex}^{(n)}(x, y) = H_{ex}^{(0)}(x - s_n, y)$$
$$H_{ey}^{(n)}(x, y) = H_{ey}^{(0)}(x - s_n, y). \qquad (n = 1, 2, 3, \ldots) \qquad (12)$$

The total field of the array is obtained by summing the contributions from all the elements,

$$H_{ex}(x, y) = \sum_{n=0}^{N_e - 1} H_{ex}^{(0)}(x - s_n, y), \qquad (13)$$

$$H_{ey}(x, y) = \sum_{n=0}^{N_e - 1} H_{ey}^{(0)}(x - s_n, y). \qquad (14)$$

The force components are given by[13,16,17]

$$F_{mx}(x, y) = \mu_0 V_c (\chi_c - \chi_f) \left\{ \left( \sum_{n=0}^{N_e-1} H_{ex}^{(0)}(x - s_n, y) \right) \left( \sum_{n=0}^{N_e-1} \frac{\partial H_{ex}^{(0)}(x - s_n, y)}{\partial x} \right) \right. $$
$$\left. + \left( H_{bias,y} + \sum_{n=0}^{N_e-1} H_{ey}^{(0)}(x - s_n, y) \right) \left( \sum_{n=0}^{N_e-1} \frac{\partial H_{ex}^{(0)}(x - s_n, y)}{\partial y} \right) \right\}, \qquad (15)$$

and

$$F_{my}(x, y) = \mu_0 V_c (\chi_c - \chi_f) \left\{ \left( \sum_{n=0}^{N_e-1} H_{ex}^{(0)}(x - s_n, y) \right) \left( \sum_{n=0}^{N_e-1} \frac{\partial H_{ey}^{(0)}(x - s_n, y)}{\partial x} \right) \right. $$
$$\left. + \left( H_{bias,y} + \sum_{n=0}^{N_e-1} H_{ey}^{(0)}(x - s_n, y) \right) \left( \sum_{n=0}^{N_e-1} \frac{\partial H_{ey}^{(0)}(x - s_n, y)}{\partial y} \right) \right\}. \qquad (16)$$

In Eqs. (15) and (16) we have assume that the bias field is constant and in the y-direction.

**B. Fluidic force**

The fluidic force is predicted using Stokes' law for the drag on a sphere in a viscous fluid. The blood flow in the microchannel is assumed to be laminar, and since the diameters of WBCs and RBCs are small relative to the channel height, we assume that the fluid velocity is relatively constant across the cells. We use Eq. (3) to estimate the drag force at a given time using the particle velocity at that time, and the fluid velocity at the position of the cell at that time. Let $h_c$ and $w_c$ denote the half-height and half-width of its rectangular cross section (Fig. 2a). We assume fully developed laminar flow parallel to the x-axis and obtain

$$v_f(y) = \frac{3\overline{v}_f}{2}\left[1 - \left(\frac{y-(h+h_c)}{h_c}\right)^2\right], \tag{17}$$

where $\overline{v}_f$ is the average flow velocity in the channel.[16,17] Note that the distance y in Eq. (17) is taken with respect the center of the magnetic elements (Fig. 2b). We substitute Eq. (17) into Eq. (3) and obtain the fluidic force components

$$\mathbf{F}_{fx} = -6\pi\eta R_{c,hyd}\left[v_{c,x} - \frac{3\overline{v}_f}{2}\left[1 - \left(\frac{y-(h+h_c)}{h_c}\right)^2\right]\right], \tag{18}$$

and

$$\mathbf{F}_{fy} = -6\pi\eta R_{c,hyd}v_{c,y}. \tag{19}$$

We use these in the equations of motion below.

**C. Blood cell properties**

The magnetic properties of WBCs, RBCs and plasma are needed to complete the model. White blood cells comprise five different kinds of cells that are classified into two groups: agranulocytes (lymphocyte and monocyte), and granulocytes (neutrophil, eosinophil and basophil).[10,20] These cells have different sizes, with diameters that range from 6 μm to 15 μm. We account for the different types of WBCs by using the following average cell properties: $\rho_{wbc} = 1070$ kg/m$^3$, $R_{wbc} = 5$ μm, and $V_{wbc} = 524$ μm$^3$.[20] White blood cells exhibit a diamagnetic behavior in plasma, but their magnetic susceptibility is not well known.[10] Furthermore, the different types of WBCs may have different values of susceptibility, and these decrease with time. In order to determine the feasibility of WBC separation we use an estimate for WBC susceptibility as suggested by Takayasu et al., specifically we use the susceptibility of water $\chi_{wbc} = \chi_{H2O} = -9.2 \times 10^{-6}$ (SI).[10] While this value should provide a conservative estimate of the magnetic force on a WBC, we study WBC separation below using a range of values ($-9.2 \times 10^{-6} \leq \chi_{wbc} < \chi_{plasma}$) in order to determine the impact of this parameter on system performance.

Red blood cells when unperturbed, have a well-defined biconcave discoid shape with a diameter of 8.5 ± 0.4 μm and a thickness of 2.3 ± 0.1 μm. These cells account for approximately 99% of the particulate matter in blood, and the percentage by volume (hematocrit) of packed red blood cells in a given sample of blood, is normally 40-45%. For red blood cells, we use $R_{rbc} = 3.84$ μm (hydrodynamic radius), $V_{rbc} = 88.4$ μm$^3$, and $\rho_{rbc} = 1100$ kg/m$^3$.[21] The susceptibility of a RBC depends on the oxygenation of its hemoglobin. We use $\chi_{rbc,oxy} = -9.22 \times 10^{-6}$ (SI) and $\chi_{rbc,deoxy} = -3.9 \times 10^{-6}$ (SI) for oxygenated and deoxygenated red blood cells, respectively.[9,10,21] Finally, the transport fluid is blood plasma which has the following properties: $\eta = 0.001$ kg/s, $\rho_f = 1000$ kg/m$^3$ and $\chi_f = -7.7 \times 10^{-6}$ (SI).[9,10,21]

**D. Equations of motion**

The equations of motion for blood cell transport through the microsystem can be written in component form by substituting Eqs. (15), (16), (18) and (19) into Eq. (1),

$$m_c \frac{dv_{c,x}}{dt} = F_{mx}(x_c, y_c) - 6\pi\eta R_{c,hyd}\left[v_{c,x} - \frac{3\bar{v}_f}{2}\left[1 - \left(\frac{y_c - (h + h_c)}{h_c}\right)^2\right]\right] + V_c(\rho_c - \rho_f)g, \quad (20)$$

$$m_c \frac{dv_{c,y}}{dt} = F_{my}(x_c, y_c) - 6\pi\eta R_{c,hyd} v_{c,y}, \quad (21)$$

$$v_{c,x}(t) = \frac{dx_c}{dt}, \qquad v_{c,y}(t) = \frac{dy_c}{dt}. \quad (22)$$

Equations (20) - (22) constitute a coupled system of first-order ordinary differential equations that are solved subject to initial conditions for $x_c(0)$, $y_c(0)$, $v_{c,x}(0)$, and $v_{c,y}(0)$. These equations can be solved numerically using numerical techniques such as the Runge-Kutta method.

## II. Results

We use Eqs. (20) - (22) to study blood cell transport through the microsystem. We start with a specific configuration in which the microchannel is 120 μm high ($h_c = 60 \mu m$), 1 mm wide and 35 mm long. There are 80 identical permalloy elements (78% Ni 22% Fe, page 43 in reference 18) embedded transverse and adjacent to the microchannel. Each element has a height and width of 200 μm, and the elements are spaced 200 μm apart (edge to edge). Thus, w = h = 100 μm in Eqs. (10) and (11), and these elements have an aspect ratio $p = h/w = 1$. We use Eqs. (7) and (8) to determine the demagnetization factor for these elements, which turns out to be $N_d = 0.456$ (p 191, Table A.2 in reference 19). The 80 elements span a distance of 31.8 mm along the channel. The bias field is 5000 Gauss ($H_{bias} = 3.9 \times 10^5$ A/m), which is sufficient to magnetically saturate the elements.

It instructive to examine the magnetic force provided by the elements. To this end, we use Eqs. (15) and (16) to predict the magnetic force on deoxygenated RBCs in the microchannel above an isolated array of three magnetized elements with the dimensions and spacing as defined above. The force components $F_{mx}$ and $F_{my}$ across the microchannel are shown in Figs 3a and 3b, respectively. These components are computed over the dotted rectangular region show in Fig. 3c, which extends from 20 μm to 100 μm above the elements, and from -100 μm to the left of the array to + 100 μm to the right of the array. Note that $F_{mx}$, which is in the flow direction, changes sign across each element; while $F_{my}$, which is responsible for cell separation, also changes sign across each element, but acts predominantly downward above each element. Thus, the magnetic elements will attract deoxygenated RBCs, but repel WBCs and oxygenated RBCs. Also not that $F_{my}$ is on the order of a pico-Newton near the elements, but decreases rapidly with distance from the elements. By comparison, the gravitational force $\mathbf{F}_g = V_c(\rho_c - \rho_f)\mathbf{g}$ (including buoyancy) on a RBC is approximately 0.09 pN, which is on the same order as, or greater than, the magnetic separation force over extended regions of the microchannel. Similarly, the flow-directed fluidic force on a RBC is even larger, (e.g. approximately 7.2 pN for an average flow velocity of $\bar{v}_f = 0.1$ mm/s ). Thus, it is advantageous to orient the microsystem with the gravitational and flow-directed fluidic force orthogonal to the direction of separation so that they do not resist it.

We now study blood cell transport. A blood sample containing WBCs and deoxygenated RBCs is introduced into the inlet. We assume that blood cells enter the microchannel to the left of the first element at $x(0) = -4w$. We choose different initial heights for the cells to determine the impact of this variable on cell separation. Specifically, we choose y(0) = 115 μm, 130 μm, …, 205 μm. The top of the microchannel is 120 μm above the elements at y = 220 μm. The average fluid velocity is $\bar{v}_f = 0.2$ mm/s, and the cells enter the channel with a velocity that correlates with their initial height in the channel (laminar flow). We use Eqs. (20) - (22) to predict WBC and RBC trajectories, which are shown in Fig. 4a and b, respectively. The trajectory profiles are slightly irregular due to the spatial variation of the magnetic force.[13,16,17] Note that the WBCs and deoxygenated RBCs separate well before they reach the end of the array. Specifically, all the WBCs move to the top of the channel, while all the deoxygenated RBCs move to the bottom. The separation times for the WBCs and RBCs (i.e. the time it takes for all of the cells to reach their respective ends of the microchannel) are 95s and 120s, respectively. As a consequence of this separation, WBC-rich plasma and deoxygenated RBC-rich plasma will exit the microchannel at the respective side outlets, while cell-depleted plasma will exit through the central outlet as shown in Fig. 1b.

Next, we study cell separation as a function of the dimensions of the magnetic elements. We repeat the analysis above but vary the dimensions and spacing of the elements with all other variables held constant. We perform two simulations with the dimensions (height, width and spacing) of the elements set to 300 μm and 400 μm, respectively. The WBC and RBC trajectories for these cases are shown in Figs. 4 and 5, respectively. Notice that cell separation occurs over a shorter distance (faster) as the size and spacing of the magnetic elements increase relative to the dimensions of the channel.

Lastly, we study WBC separation as a function of WBC susceptibility. Note that in our previous simulations we used $\chi_{wbc} = -9.2 \times 10^{-6}$ (SI), which we believe provides a conservative estimate for the magnetic force, and hence separation. However, since $\chi_{wbc}$ is not well known, it is instructive to study the feasibility of WBC separation for less negative values of $\chi_{wbc}$, i.e. in the range $-9.2 \times 10^{-6} < \chi_{wbc} < \chi_{plasma}$. Such values render a weaker magnetic force than the nominal value, and hence weaker separation. We perform simulations using $\chi_{wbc} = -8.0 \times 10^{-6}, -8.15 \times 10^{-6}, -8.45 \times 10^{-6}, \ldots, -9.20 \times 10^{-6}$ (SI). These susceptibilities yield magnetic forces that are respectively 20%, 30%, 40%,…, and 100% of the nominal force obtained using the nominal value $\chi_{wbc} = \chi_{H2O}$. We predict the WBC separation time (i.e., the time it takes for a WBC to traverse the height of the microchannel) for the range of $\chi_{wbc}$. We set the height, width and spacing of the magnetic elements to 400 μm, and set all other parameters as in our simulations above. In Fig. 7, we plot the separation time as a function of the percentage of force that each $\chi_{wbc}$ value corresponds to, i.e., $\left[(\chi_{wbc} - \chi_{plasma})/(\chi_{H2O} - \chi_{plasma})\right] \times 100 = 20, 40,…, 100$. The WBC trajectories for two of the cases: $\chi_{wbc} = -8.0 \times 10^{-6}$ and $-8.45 \times 10^{-6}$, which correspond to 20% and 50% of the nominal magnetic force, are shown in Fig. 8. From these plots we find that a weaker magnetic force renders a longer separation time as expected. However, the analysis indicates that WBC separation is still feasible even when the magnetic force is only 20% of our assumed conservative value. Furthermore, even if the magnetic coupling to the WBCs were weaker than our simulated lower bound ($\chi_{wbc} = -8.0 \times 10^{-6}$), there are several other variables that can be adjusted to compensate for this to ensure viable WBC separation. Specifically, one could increase the number, size, and spacing of the elements, or decrease the height of the microchannel, or reduce the flow rate to enhance separation for weakly coupled cells.

## IV. Conclusion

We have presented a method and model for the direct and continuous separation of deoxygenated whole blood into three components: deoxygenated RBC-rich plasma, WBC-rich plasma, and cell-depleted plasma. The method can be implemented in a passive magnetophoretic microsystem that enables rapid processing of small blood samples. The microsystem is compact, efficient and can be fabricated using established methods.[12,22-24] As such, it holds substantial potential for point-of-service medical diagnostic applications. The microsystem is also suitable for a broad range of applications that involve the manipulation, immobilization or filtering of weakly magnetic micro or nanoparticles.

# Figure Captions

FIG. 1. Magnetophoretic microsystem: (a) microsystem with bias field structure; (b) cross-section of microsystem showing magnetic elements, microchannel and outlets; and (c) magnified view of microfluidic channel showing the bias field, magnetic elements, and forces on red and white blood cells (RBCs and WBCs).

FIG. 2. Magnetophoretic microsystem: (a) microfluidic channel, and (b) cross section of microsystem showing array of magnetized elements.

FIG. 3. Magnetic force across the microchannel above three magnetized elements: (a) surface plot of $F_{mx}$, (b) surface plot of $F_{my}$., (c) area (dotted line) over which the magnetic force is computed.

FIG. 4. Blood cell trajectories above magnetized elements; height, width and spacing of elements = 200 μm (upper half of elements shown for reference): (a) WBC trajectories, (b) RBC trajectories.

FIG. 5. Blood cell trajectories above magnetized elements, height, width and spacing of elements = 300 μm (upper half of elements shown for reference): (a) WBC trajectories, (b) RBC trajectories.

FIG. 6. Blood cell trajectories above magnetized elements, height, width and spacing of elements = 400 μm (upper half of elements shown for reference): (a) WBC trajectories, (b) RBC trajectories.

FIG. 7. White blood cell separation time vs. WBC susceptibility.

FIG. 8. White blood cell trajectories above magnetized elements, height, width and spacing of elements = 400 μm (upper half of elements shown for reference): (a) $\chi_{wbc} = -8.45 \times 10^{-6}$ (SI), (b) $\chi_{wbc} = -8.0 \times 10^{-6}$ (SI).

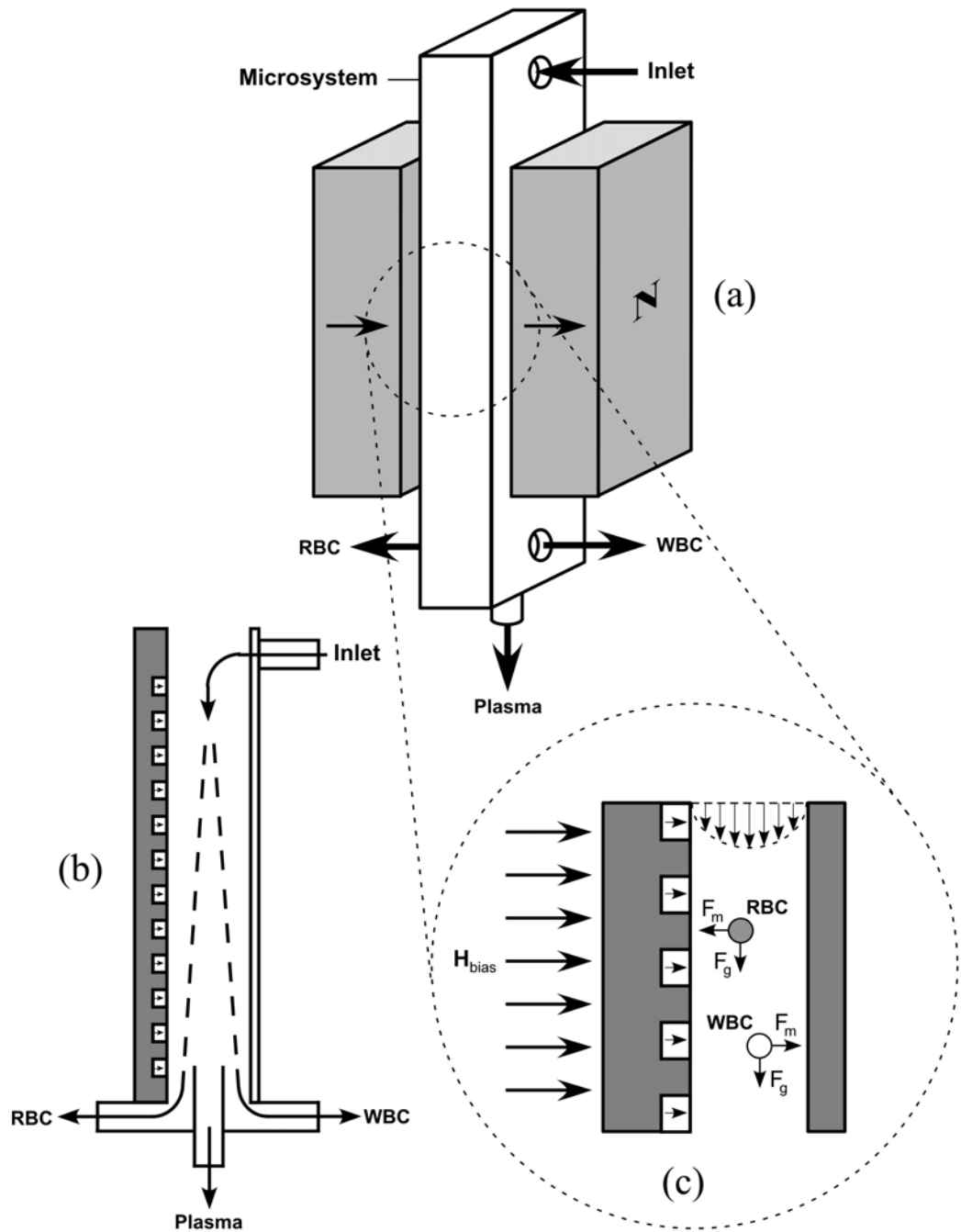

FIG. 1

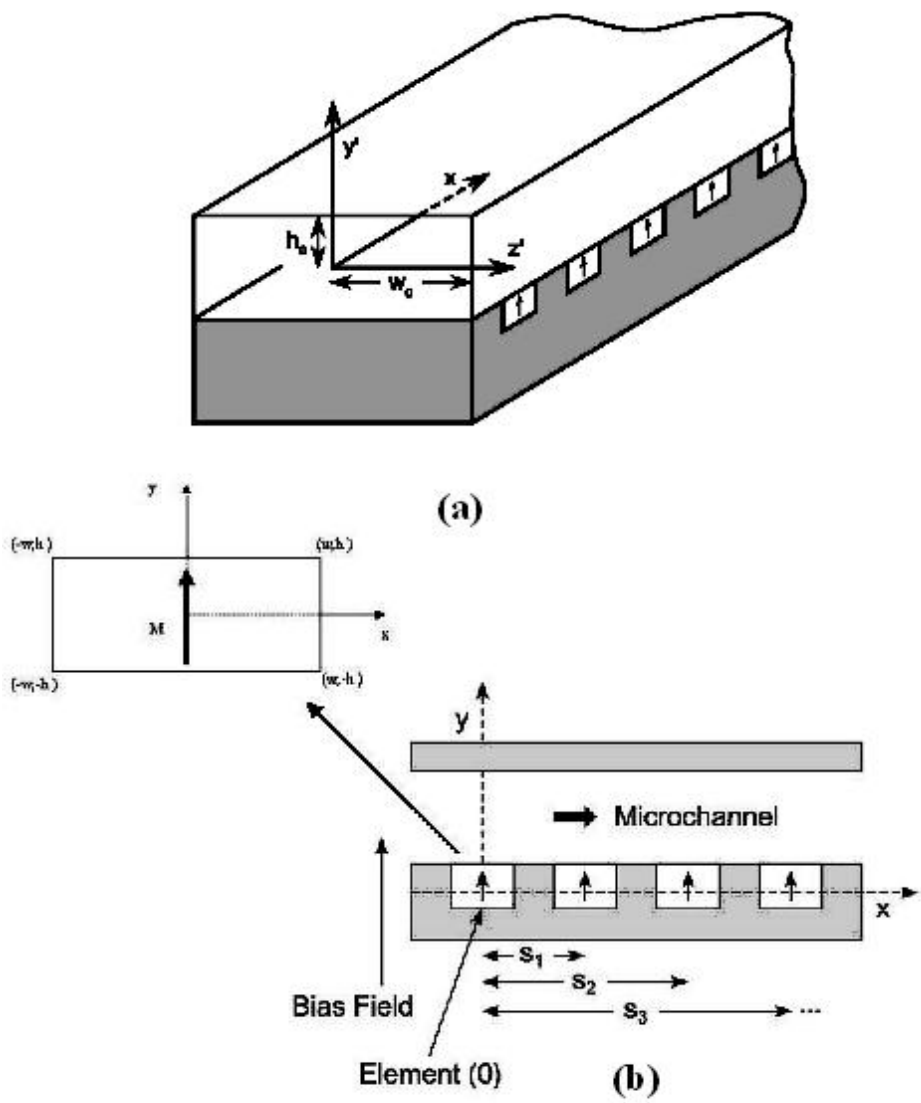

FIG. 2

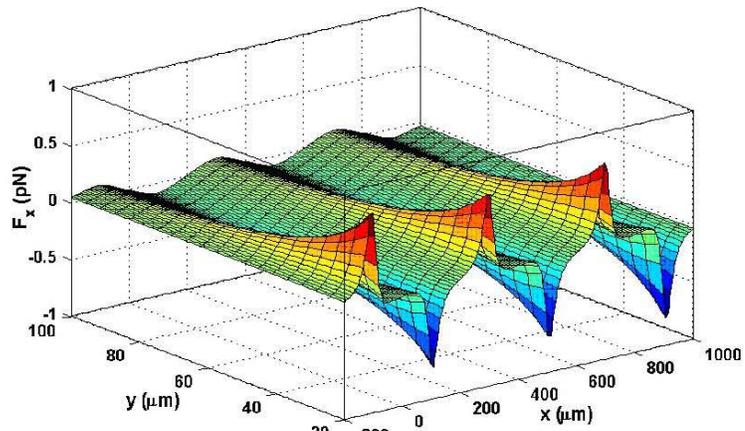

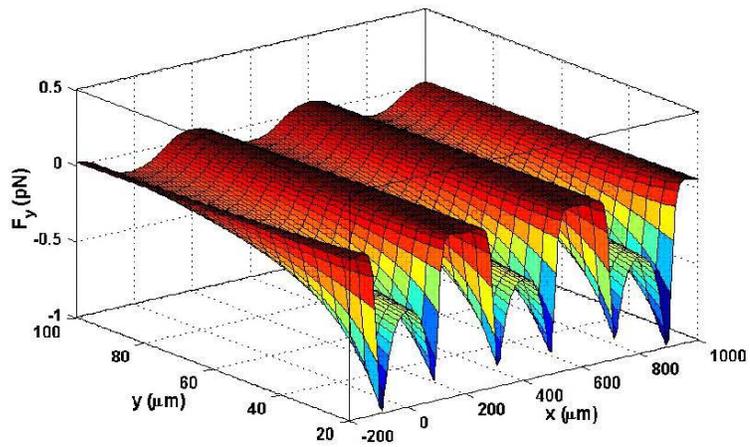

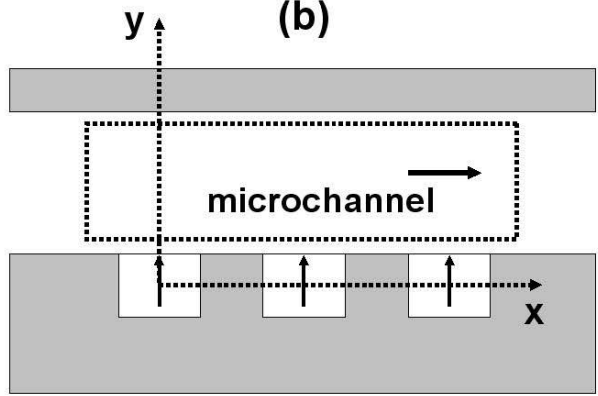

FIG. 3

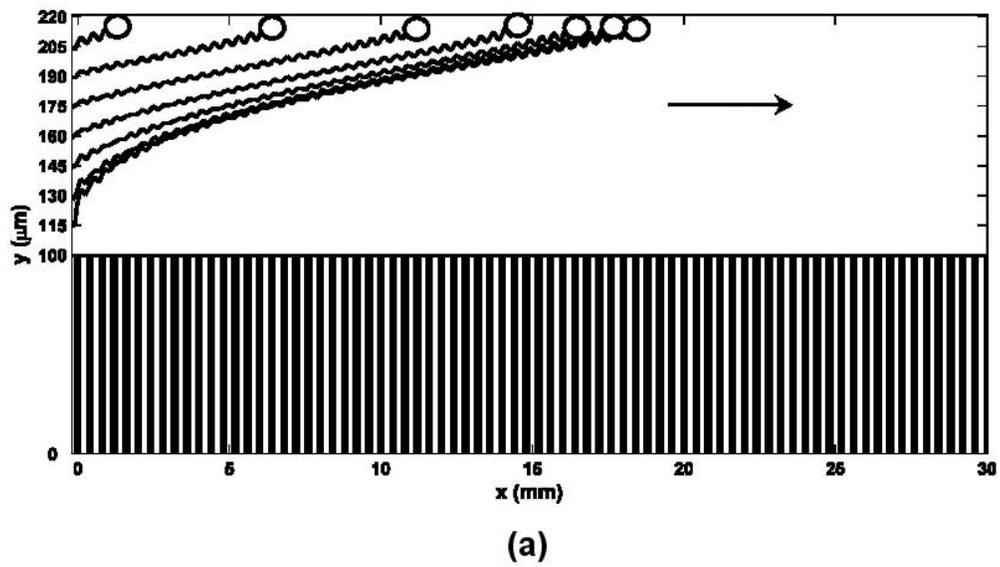

(a)

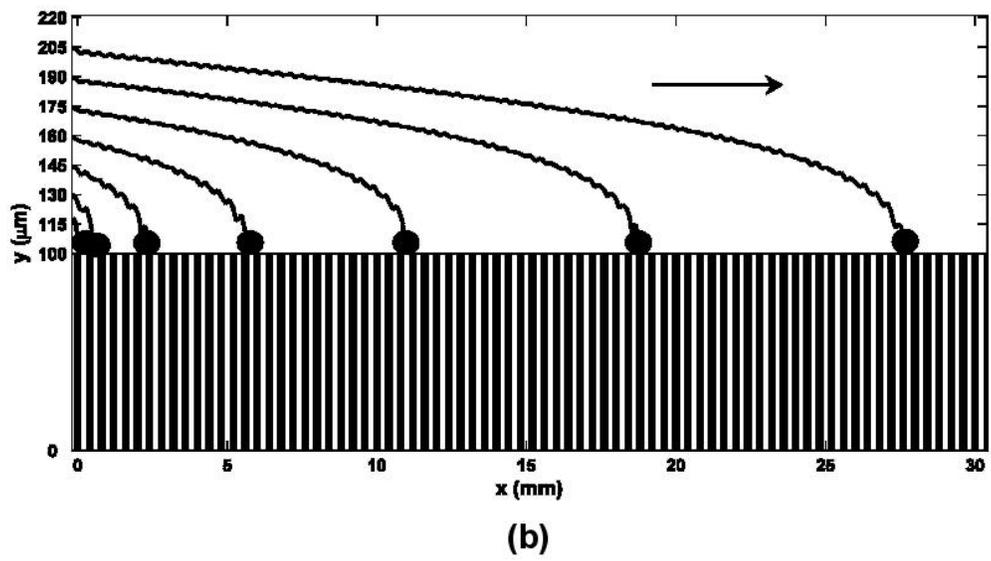

(b)

FIG. 4

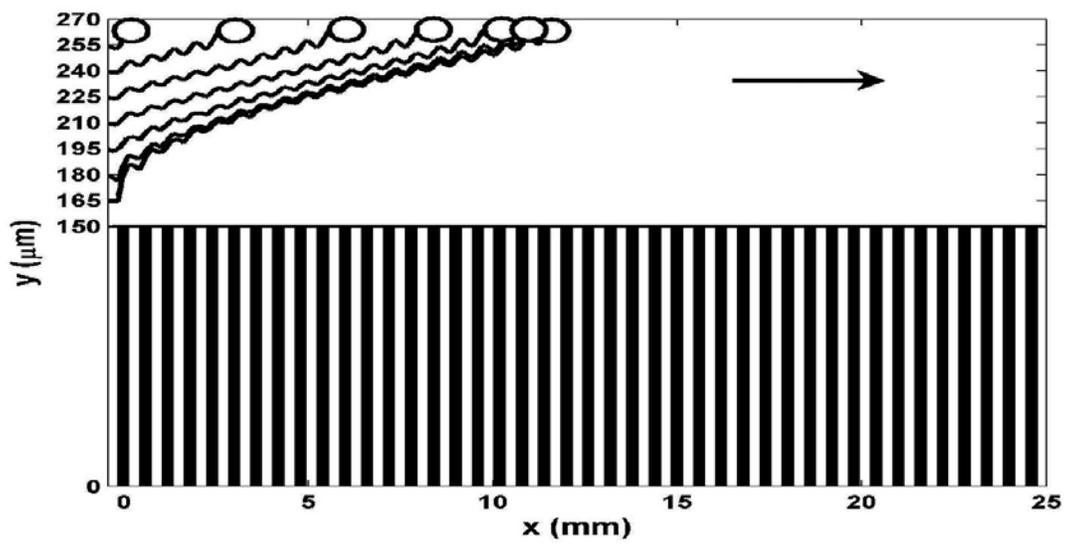

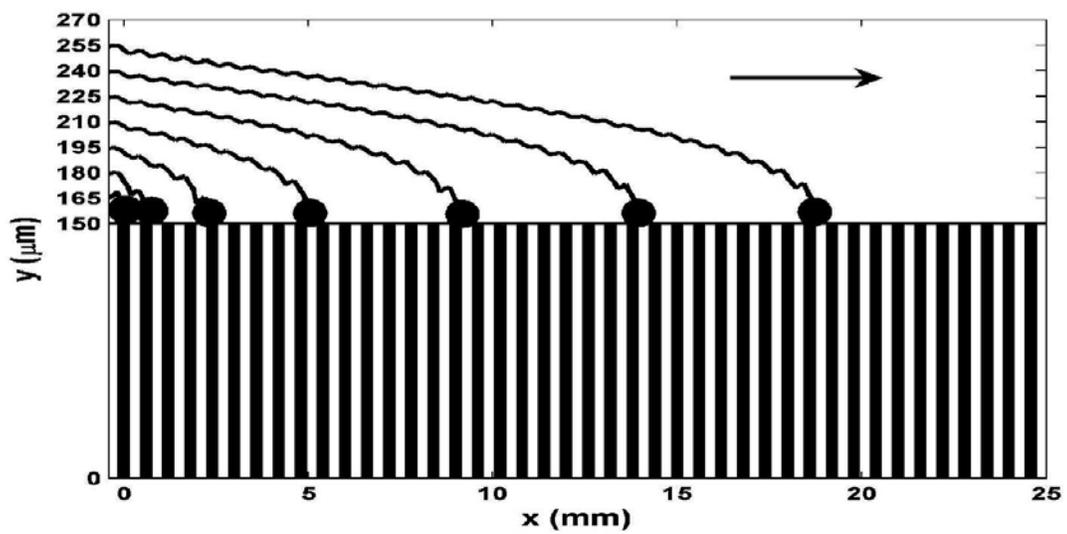

FIG. 5

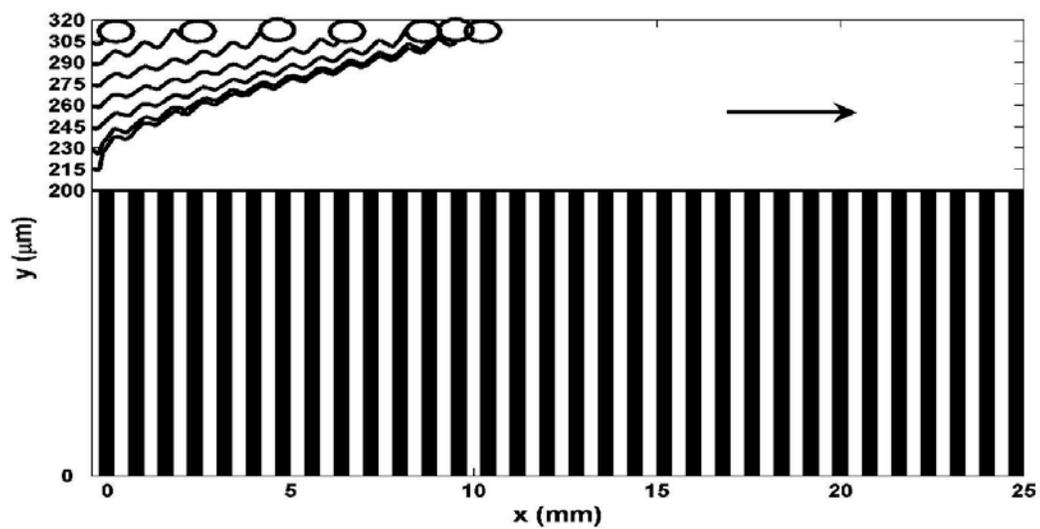

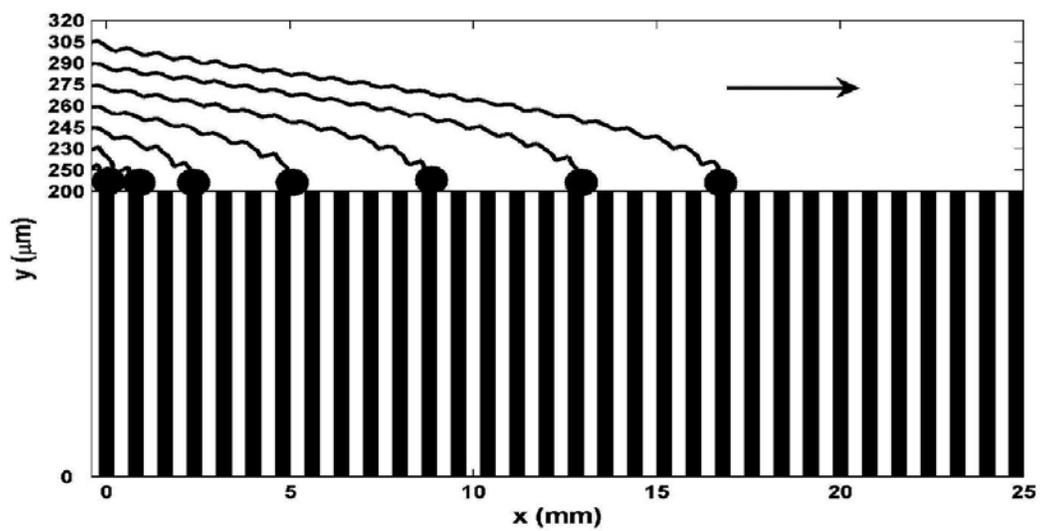

FIG. 6

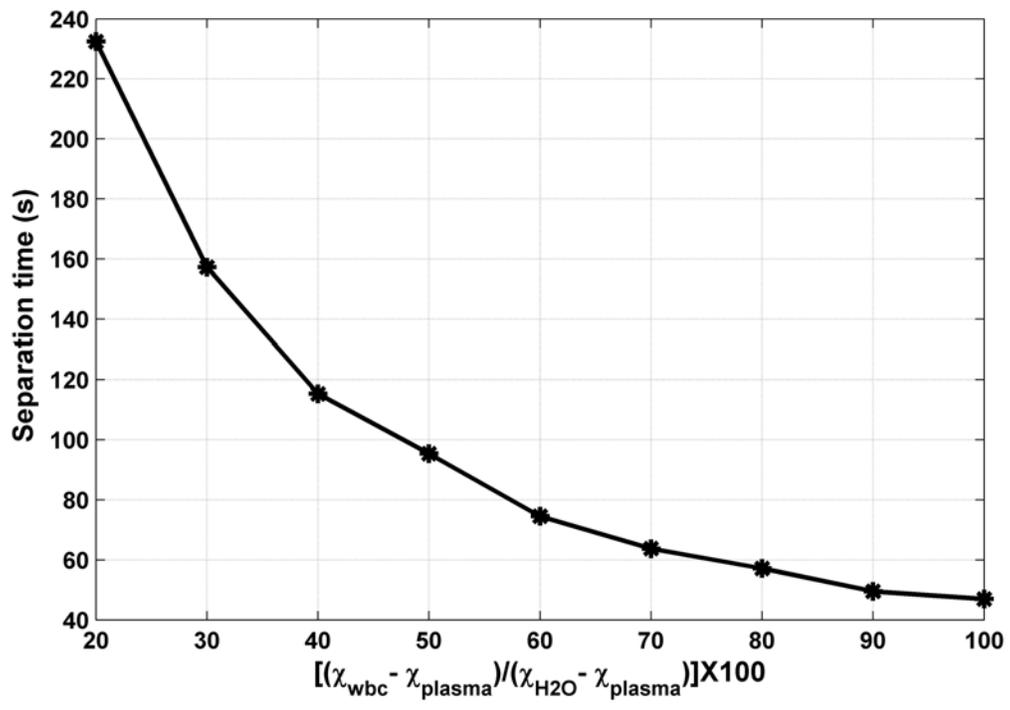

FIG. 7.

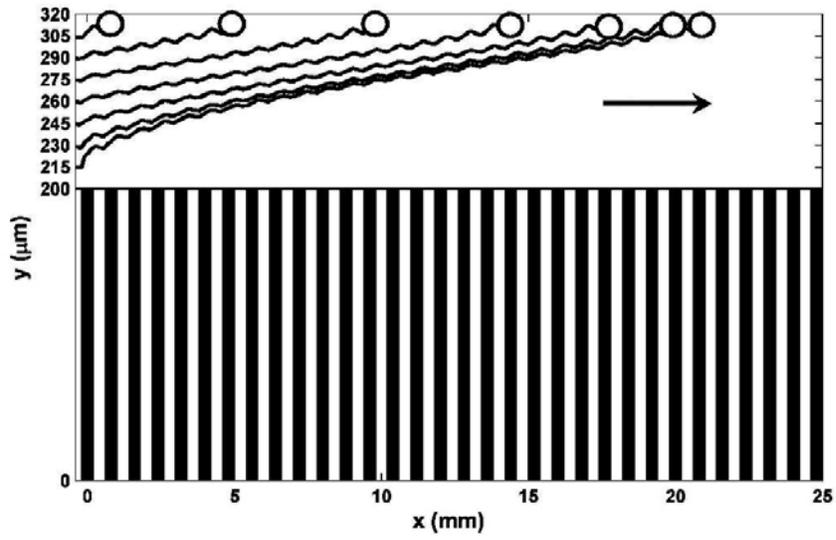

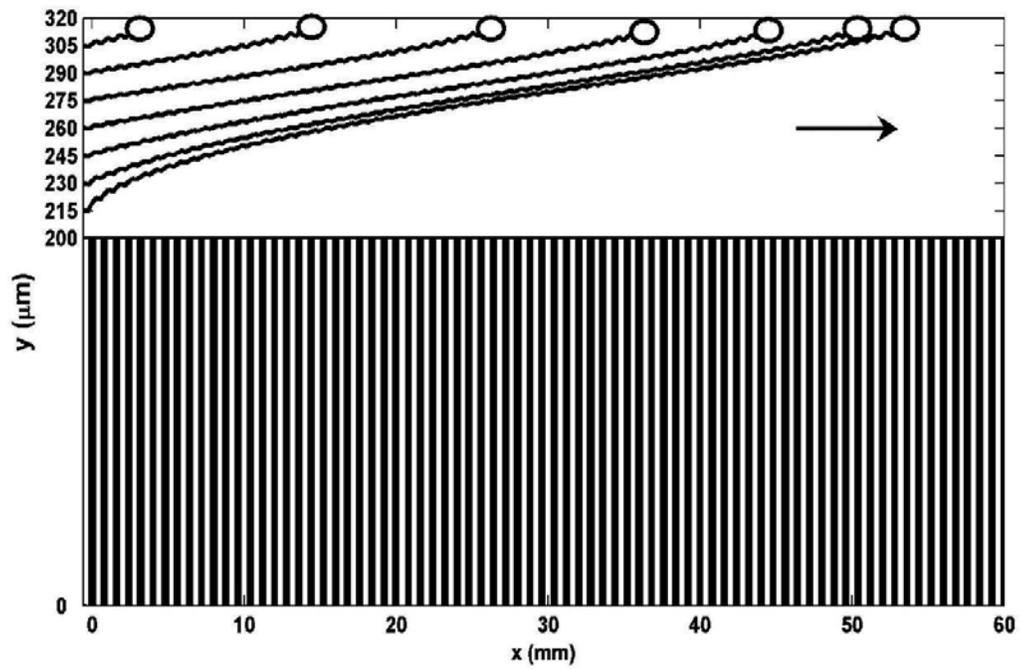

FIG. 8